\documentclass[nature,twocolumn,superscriptaddress]{revtex4-2}

\usepackage{graphicx}
\usepackage{dcolumn}
\usepackage{bm}
\usepackage{amsmath}
\usepackage{amssymb}
\usepackage{latexsym}
\usepackage{epsfig}
\usepackage{amsbsy}
\usepackage{array}
\usepackage{amssymb}
\usepackage{setspace}
\usepackage{color}
\usepackage{url}
\usepackage{hyperref}
\usepackage{multirow}
\usepackage{comment}

\begin{document}
\preprint{Nature Communications}
\title{Proton-Boron Fusion Yield Increased by Orders of Magnitude with Foam Targets}

\author{Wen-Qing Wei} \thanks{These authors contributed equally} \affiliation{Ministry of Education Key Laboratory for Nonequilibrium Synthesis and Modulation of Condensed Matter, Shaanxi Province Key Laboratory of Quantum Information and Quantum Optoelectronic Devices, School of Physics, Xi'an Jiaotong University, Xi'an 710049, China}

\author{Shi-Zheng Zhang} \thanks{These authors contributed equally} \affiliation{Ministry of Education Key Laboratory for Nonequilibrium Synthesis and Modulation of Condensed Matter, Shaanxi Province Key Laboratory of Quantum Information and Quantum Optoelectronic Devices, School of Physics, Xi'an Jiaotong University, Xi'an 710049, China}	

\author{Zhi-Gang Deng} \thanks{These authors contributed equally} \affiliation{Science and Technology on Plasma Physics Laboratory, Laser Fusion Research Center, China Academy of Engineering Physics, Mianyang 621900, China}

\author{Wei Qi} \affiliation{Science and Technology on Plasma Physics Laboratory, Laser Fusion Research Center, China Academy of Engineering Physics, Mianyang 621900, China}

\author{Hao Xu} \affiliation{Ministry of Education Key Laboratory for Nonequilibrium Synthesis and Modulation of Condensed Matter, Shaanxi Province Key Laboratory of Quantum Information and Quantum Optoelectronic Devices, School of Physics, Xi'an Jiaotong University, Xi'an 710049, China}	

\author{Li-Rong Liu} \affiliation{Ministry of Education Key Laboratory for Nonequilibrium Synthesis and Modulation of Condensed Matter, Shaanxi Province Key Laboratory of Quantum Information and Quantum Optoelectronic Devices, School of Physics, Xi'an Jiaotong University, Xi'an 710049, China}

\author{Jia-Lin Zhang} \affiliation{Ministry of Education Key Laboratory for Nonequilibrium Synthesis and Modulation of Condensed Matter, Shaanxi Province Key Laboratory of Quantum Information and Quantum Optoelectronic Devices, School of Physics, Xi'an Jiaotong University, Xi'an 710049, China}

\author{Fang-Fang Li} \affiliation{Ministry of Education Key Laboratory for Nonequilibrium Synthesis and Modulation of Condensed Matter, Shaanxi Province Key Laboratory of Quantum Information and Quantum Optoelectronic Devices, School of Physics, Xi'an Jiaotong University, Xi'an 710049, China}

\author{Xing Xu} \affiliation{Ministry of Education Key Laboratory for Nonequilibrium Synthesis and Modulation of Condensed Matter, Shaanxi Province Key Laboratory of Quantum Information and Quantum Optoelectronic Devices, School of Physics, Xi'an Jiaotong University, Xi'an 710049, China} \affiliation{Institute of Modern Physics, Chinese Academy of Sciences, Lanzhou 710049, China}

\author{Zhong-Min Hu} \affiliation{Ministry of Education Key Laboratory for Nonequilibrium Synthesis and Modulation of Condensed Matter, Shaanxi Province Key Laboratory of Quantum Information and Quantum Optoelectronic Devices, School of Physics, Xi'an Jiaotong University, Xi'an 710049, China}

\author{Ben-Zheng Chen} \affiliation{Ministry of Education Key Laboratory for Nonequilibrium Synthesis and Modulation of Condensed Matter, Shaanxi Province Key Laboratory of Quantum Information and Quantum Optoelectronic Devices, School of Physics, Xi'an Jiaotong University, Xi'an 710049, China}

\author{Bu-Bo Ma} \affiliation{Ministry of Education Key Laboratory for Nonequilibrium Synthesis and Modulation of Condensed Matter, Shaanxi Province Key Laboratory of Quantum Information and Quantum Optoelectronic Devices, School of Physics, Xi'an Jiaotong University, Xi'an 710049, China}

\author{Jian-Xing Li} \affiliation{Ministry of Education Key Laboratory for Nonequilibrium Synthesis and Modulation of Condensed Matter, Shaanxi Province Key Laboratory of Quantum Information and Quantum Optoelectronic Devices, School of Physics, Xi'an Jiaotong University, Xi'an 710049, China}

\author{Xue-Guang Ren} \affiliation{Ministry of Education Key Laboratory for Nonequilibrium Synthesis and Modulation of Condensed Matter, Shaanxi Province Key Laboratory of Quantum Information and Quantum Optoelectronic Devices, School of Physics, Xi'an Jiaotong University, Xi'an 710049, China}

\author{Zhong-Feng Xu} \affiliation{Ministry of Education Key Laboratory for Nonequilibrium Synthesis and Modulation of Condensed Matter, Shaanxi Province Key Laboratory of Quantum Information and Quantum Optoelectronic Devices, School of Physics, Xi'an Jiaotong University, Xi'an 710049, China}

\author{Dieter H. H. Hoffmann} \affiliation{Ministry of Education Key Laboratory for Nonequilibrium Synthesis and Modulation of Condensed Matter, Shaanxi Province Key Laboratory of Quantum Information and Quantum Optoelectronic Devices, School of Physics, Xi'an Jiaotong University, Xi'an 710049, China}

\author{Quan-Ping Fan} \affiliation{Science and Technology on Plasma Physics Laboratory, Laser Fusion Research Center, China Academy of Engineering Physics, Mianyang 621900, China}

\author{Wei-Wu Wang} \affiliation{Science and Technology on Plasma Physics Laboratory, Laser Fusion Research Center, China Academy of Engineering Physics, Mianyang 621900, China}

\author{Shao-Yi Wang} \affiliation{Science and Technology on Plasma Physics Laboratory, Laser Fusion Research Center, China Academy of Engineering Physics, Mianyang 621900, China}

\author{Jian Teng} \affiliation{Science and Technology on Plasma Physics Laboratory, Laser Fusion Research Center, China Academy of Engineering Physics, Mianyang 621900, China}

\author{Bo Cui} \affiliation{Science and Technology on Plasma Physics Laboratory, Laser Fusion Research Center, China Academy of Engineering Physics, Mianyang 621900, China}

\author{Feng Lu} \affiliation{Science and Technology on Plasma Physics Laboratory, Laser Fusion Research Center, China Academy of Engineering Physics, Mianyang 621900, China}

\author{Lei Yang} \affiliation{Science and Technology on Plasma Physics Laboratory, Laser Fusion Research Center, China Academy of Engineering Physics, Mianyang 621900, China}

\author{Yu-Qiu Gu} \affiliation{Science and Technology on Plasma Physics Laboratory, Laser Fusion Research Center, China Academy of Engineering Physics, Mianyang 621900, China}

\author{Zong-Qing Zhao} \affiliation{Science and Technology on Plasma Physics Laboratory, Laser Fusion Research Center, China Academy of Engineering Physics, Mianyang 621900, China}

\author{Rui Cheng} \affiliation{Institute of Modern Physics, Chinese Academy of Sciences, Lanzhou 710049, China}

\author{Zhao Wang} \affiliation{Institute of Modern Physics, Chinese Academy of Sciences, Lanzhou 710049, China}

\author{Yu Lei} \affiliation{Institute of Modern Physics, Chinese Academy of Sciences, Lanzhou 710049, China}

\author{Guo-Qing Xiao} \affiliation{Institute of Modern Physics, Chinese Academy of Sciences, Lanzhou 710049, China}

\author{Hong-Wei Zhao} \affiliation{Institute of Modern Physics, Chinese Academy of Sciences, Lanzhou 710049, China}

\author{Bing Liu}  \affiliation{Hebei Key Laboratory of Compact Fusion, Langfang 065001, China} \affiliation{ENN Science and Technology Development Co., Ltd., Langfang 065001, China}

\author{Guan-Chao Zhao} \affiliation{Hebei Key Laboratory of Compact Fusion, Langfang 065001, China} \affiliation{ENN Science and Technology Development Co., Ltd., Langfang 065001, China}

\author{Min-Sheng Liu} \affiliation{Hebei Key Laboratory of Compact Fusion, Langfang 065001, China} \affiliation{ENN Science and Technology Development Co., Ltd., Langfang 065001, China}

\author{Hua-Sheng Xie} \affiliation{Hebei Key Laboratory of Compact Fusion, Langfang 065001, China} \affiliation{ENN Science and Technology Development Co., Ltd., Langfang 065001, China}

\author{Lei-Feng Cao} \affiliation{Advanced Materials Testing Technology Research Center, Shenzhen University of Technology, Shenzhen, 518118, China}

\author{Jie-Ru Ren} \email{renjieru@xjtu.edu.cn} \affiliation{Ministry of Education Key Laboratory for Nonequilibrium Synthesis and Modulation of Condensed Matter, Shaanxi Province Key Laboratory of Quantum Information and Quantum Optoelectronic Devices, School of Physics, Xi'an Jiaotong University, Xi'an 710049, China}	

\author{Wei-Min Zhou} \email{zhouwm@caep.cn} \affiliation{Science and Technology on Plasma Physics Laboratory, Laser Fusion Research Center, China Academy of Engineering Physics, Mianyang 621900, China}
	
\author{Yong-Tao Zhao} \email{zhaoyongtao@xjtu.edu.cn} \affiliation{Ministry of Education Key Laboratory for Nonequilibrium Synthesis and Modulation of Condensed Matter, Shaanxi Province Key Laboratory of Quantum Information and Quantum Optoelectronic Devices, School of Physics, Xi'an Jiaotong University, Xi'an 710049, China}	

\date{\today}

\begin{abstract}
A novel intense beam-driven scheme for high yield of the tri-alpha reaction $\mathrm{^{11}B}(p,\alpha)2\alpha$ was investigated. We used a foam target made of cellulose triacetate (TAC, $\mathrm{C_9H_{16}O_8}$) doped with boron. It was then heated volumetrically by soft X-ray radiation from a laser heated hohlraum and turned into a homogenous, and long living plasma. We employed a picosecond laser pulse to generate a high-intensity energetic proton beam via the well-known Target Normal Sheath Acceleration (TNSA) mechanism. We observed up to $10^{10}$/sr $\alpha$ particles per laser shot. This constitutes presently the highest yield value normalized to the laser energy on target. The measured fusion yield per proton exceeds the classical expectation of beam-target reactions by up to four orders of magnitude under high proton intensities. This enhancement is attributed to the strong electric fields and non-equilibrium thermonuclear fusion reactions as a result of the new method. Our approach shows opportunities to pursue ignition of aneutronic fusion.

\end{abstract}

\maketitle

\section{Introduction}

The quest for fusion energy is currently dominated by the deuterium-tritium (DT) reaction. Proton-boron (p$^{11}$B) fusion rarely caught the eye, and was deemed to be insignificant. However,  it has recently become the focus of interest for many research groups since the main reaction channel offers the possibility of largely aneutronic fusion. Reactor concepts based on this reaction are less complicated with respect to activation issues of reactor material and neutron damage to high-tech equipment and superconductive material close to the reactor chamber. The energy release in form of charged particles also offers the possibility for direct conversion into electric power. Tritium breeding, which is an indispensable requirement for DT-fusion energy has not yet been demonstrated on a technical scale and tritium handling requires high safety standards. The fusion fuel constituents, boron and hydrogen, as well as the reaction products are not radioactive, thus this concept is environmentally friendly and  has the chance of being economically competitive with existing power sources. 

Recently, the ignition of a burning DT-fusion plasma in the laboratory has been demonstrated \cite{Shawareb2022,Zylstra2022,Kritcher2022}. The maximum Proton-Boron fusion cross section is at 600~keV as compared to 60~keV for DT fusion. However, the high temperature and the radiation loss due to the Z=5 boron component of the fuel pose significant problems. Currently it seems impossible to operate a proton-boron reactor with a plasma in thermal equilibrium. Many groups worldwide are investigating these issues \cite{Kurilenkov2021,Eliezer2020a,Hora2017r,Hora2017n}. Since 2005, an increasing yield of $\alpha$ particles has been observed in experiments with laser-generated proton beams and boron targets \cite{Belyaev2005,Kimura2009,Labaune2013,Picciotto2014,Margarone2015,Baccou2015n,Giuffrida2020,Margarone2020g,Bonvalet2021,Margarone2022},
and recently $\alpha$ particles from p$^{11}$B fusion have been measured in a magnetically confined plasma for the first time \cite{Magee2023}. Here, we report the highest yield yet, which constitutes a breakthrough and opens up a pathway to p$^{11}$B fusion. 

High intensity short pulse lasers are an ideal tool to induce non-thermal conditions. In the past two decades, there have been rapid advances in ultra-intense laser techniques \cite{Danson2019,Yoon2019a,Yoon2021b}, accompanied by the development of relatively low-cost and efficient laser-driven plasma accelerators \cite{Macchi2013,Daido2012}. Nowadays, intense proton beams and even heavy-ion beams with high energy can be robustly generated \cite{Higginson2018}. Due to the intense proton beam strong transient electromagnetic fields are induced that modify the energy deposition characteristic \cite{Ren2020}. However, the concurrence and interrelation of these effects are not fully elucidated. To assess if these schemes can be the basis of p$^{11}$B driven fusion energy, more experiments and quantitative analysis of the p$^{11}$B fusion reaction in plasma are needed. 

The experiment was carried out at the XG-III laser facility  located at the Laser  Fusion Research Center in Mianyang \cite{Zhu2017t}. We introduced a novel beam-driven scheme for the p$^{11}$B fusion reaction in homogenous plasma.  In this configuration an ultra-intense picosecond laser pulse generates an intense spectrum of high-intensity energetic protons up to 14~MeV, while soft X-ray hohlraum radiation volumetrically heats a boron-doped foam target. The $\alpha$ reaction yields using foam target are already up to three orders of magnitude higher than the classical beam-target interaction scheme. Turning the foam into high-temperature plasma state adds another order of magnitude. The reaction rate depends significantly on the proton beam intensity. Furthermore, the normalized $\alpha$ yield per joule of laser energy is currently the highest. In this sense, our results represent a scientific breakthrough.

\section{Method}

\begin{figure}
\centering
\begin{minipage}[b]{0.5\textwidth}
\centering
\includegraphics[width=3.4in]{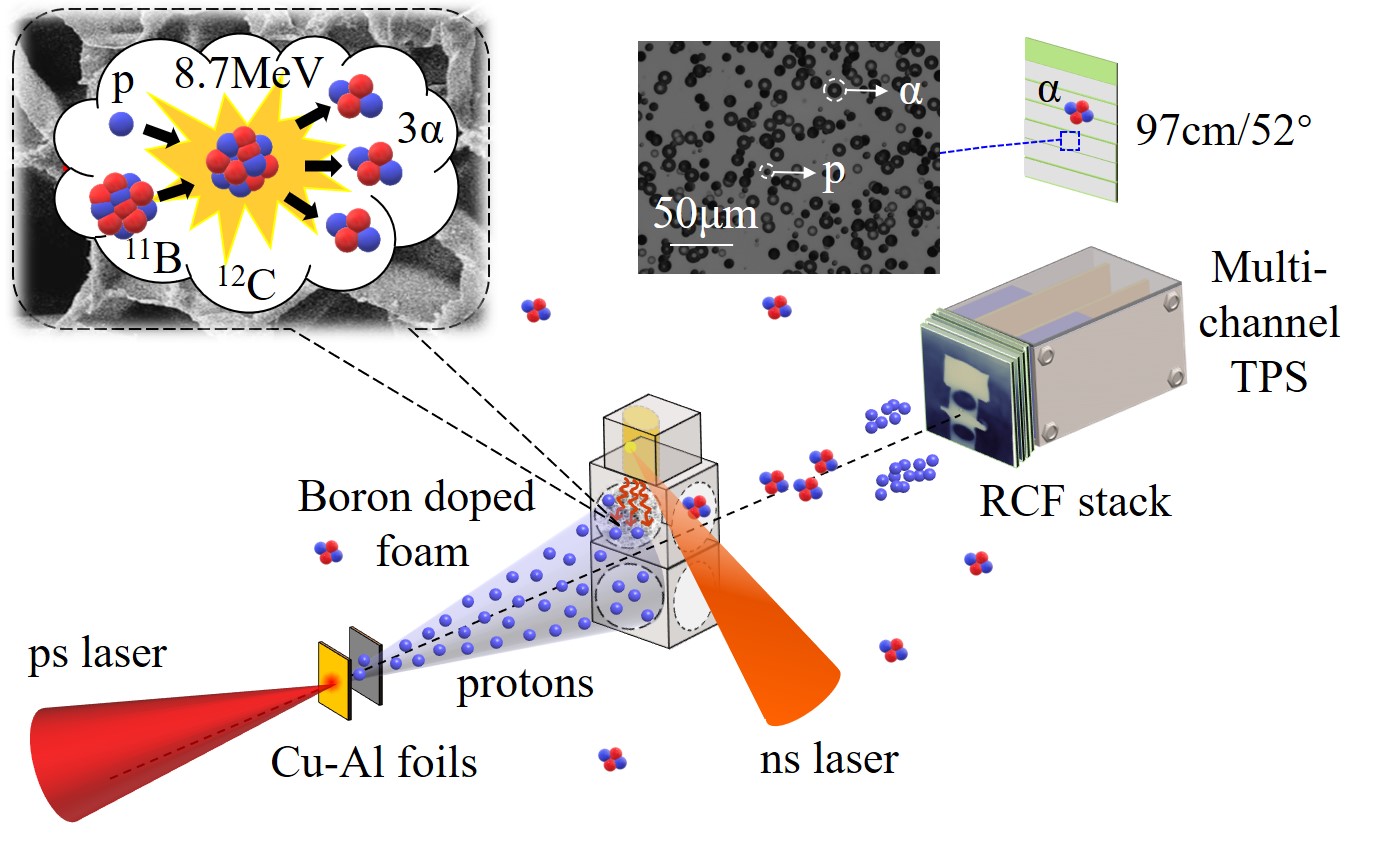}
\end{minipage}
\caption{Layout of the experiment. A ps laser is focused onto a copper foil to generate high-intensity energetic proton beam. Such protons irradiate a boron doped plasma target induced by a ns laser-driven soft X-rays to trigger the $\mathrm{p^{11}B}$ fusion reaction (see the inset). The target consists of boron doped TAC foam attached to the lower side of the cylindrical gold hohlraum. The protons and $\alpha$ particles are measured by using RCF stack, a multi-channel Thomson parabola spectrometer (TPS) and CR39 detectors covered by aluminum (Al) foils with different thicknesses, respectively. 
\label{fig1}}
\end{figure}

The experimental setup, except for the boron doping of the target and the quasi-monoenergetic incident protons, is described in detail in Ref.\cite{Ren2020}. For convenience, we repeat it here. 
As shown in Fig.~\ref{fig1}, a $p$-polarized short laser pulse ($\lambda_0=1053~$nm and $\tau_0=0.8\pm0.1~$ps) was focused onto a $10~\mathrm{\mu m}$-thick copper foil at an incident angle of $10^{\circ}$ to generate the high-intensity energetic proton beam via the TNSA mechanism \cite{Macchi2013}. The laser energy on the target was $120\pm20~$J with a focal spot of about $10~\mathrm{\mu m}$ in full width at half maximum (FWHM) and $30\%$ energy contained within. This corresponds to a peak intensity of $(2.3\pm0.3)\times10^{20}~$W/cm$^2$.  Through an 0.7~mm diameter entrance hole, a long laser pulse ($\lambda_0=527~$nm, $\tau_0=2~$ns, and $E=130\pm20~$J) was incident onto the inner wall of a cylindrical hohlraum ($\phi=1$~mm, height=1.8~mm) which had a $15~\mathrm{\mu m}$ thick gold-coating. The gold hohlraum wall was irradiated 8 ns before the arrival of the ps pulse. The generated soft X-ray flux subsequently irradiated and volumetrically heated a cellulose triacetate (TAC, $\mathrm{C_9H_{16}O_8}$) foam target, which was doped with $20\%$ natural boron. The target parameters are: diameter $\phi=1~$mm and density $\rho=2~$mg/cm$^3$.  The foam target was placed 11.5~mm downstream of the copper foil where the proton beam was generated. A $10~\mathrm{\mu m}$-thick aluminum (Al) foil covered the backside of the copper foil holder to shield stray intensity of the ns pulse and to filter part of the low-energy ion beams. The proton beam irradiated the foam target to trigger the $\mathrm{p^{11}B}$ fusion reaction, as well as through a reference empty hole, which is symmetrically located at the bottom. For several shots the foam targets were not heated by the ns laser-driven X-rays. We marked them as cold foam cases.

The hydrodynamic response to soft X-ray heating and the plasma parameters of the foam target  are well investigated and characterized \cite{Rosmej2015,Rosmej2011,Faik2014}. The sponge like structure of the target material expands, while keeping the volume and density nearly constant for more than 10~ns. This results in a homogenization of the target material. Hence, a large-scale, ns-long-living, homogeneous and quasi-static boron doped plasma is generated. In order to characterize the plasma state, emission spectra of the gold hohlraum and the foam target were measured by a transmission and a flat-field grating spectrometer (TGS and FGS) coupled to a calibrated image plate (IP, type BAS-TR), respectively \cite{Mabb2022,SM2022}.  The hohlraum radiation agrees well with a black body radiation at 19~eV. The temperature of the plasma target is about 17~eV obtained from the Boltzmann-plot method. We used the collisional radiative code FLYCHK \cite{Chung2005} to obtain the ionization degree of the boron doped CHO foam. It is around $\mathrm{C^{3.8+}H^{0.98+}O^{4.5+}B^{2.9+}}$ and the corresponding free electron density is $n_e\approx4\times10^{20}~\mathrm{cm^{-3}}$, and the boron density is $n_{^{11}\mathrm{B}}\approx1.8\times10^{19}~\mathrm{cm^{-3}}$.

The spatial-intensity distribution of the proton beam was diagnosed in discrete energy slices using a calibrated radiochromic film (RCF, type HD-V2) stack with a 4~mm-wide slit at the centre, positioned 11.4~cm downstream from the copper foil. A $15~\mathrm{\mu m}$-thick Al foil covered the front of the RCF stack to shield it from the low-energy protons ($E_p\leq1~$MeV) and carbons ions ($E_\mathrm{C}\leq18~$MeV), target debris and direct laser irradiation. The total dose in each film, allows to extract the energy spectra of the proton beam before and after the foam target \cite{Weiwq2018,Chensn2016,Devic2016}. In addition, a multi-channel Thomson Parabola Spectrometer (TPS) with an IP detector was used to measure the energy spectra of the ion beams next to the stack with an acceptance angle of $1.4\times10^{-6}~$sr. The incident total proton number and energies are obtained by combining the TPS and RCF stack results \cite{SM2022}. Track detectors CR39 \cite{Jeongtw2017} covered by Al foils of thickness from $5~\mathrm{\mu m}$ to $40~\mathrm{\mu m}$ ($2-17~\mathrm{\mu m}$ in some cases) were utilized to measure the absolute number and the energy spectra of $\alpha$ particles. The CR39 detector covered by Al foils of thickness between 5 and 20~$\mathrm{\mu m}$ was calibrated with a standard $^{241}$Am source emitting $\alpha$ particles mainly at 5.49~MeV \cite{SM2022}. Then the track diameter of the $\alpha$ particles within achieved energies was between 9 and $12~\mathrm{\mu m}$ after 3~h etching in a 6~mol/L NaOH solution with a constant temperature of $80^{\circ}\mathrm{C}$ \cite{Baccou2015,Zakifm2007,Damien1999}. These detectors were placed at different directions (with respect to the copper-foil target normal) around the foam target at different distances.  

\begin{figure}
\centering
\begin{minipage}[b]{0.5\textwidth}
\centering
\includegraphics[width=3.4in]{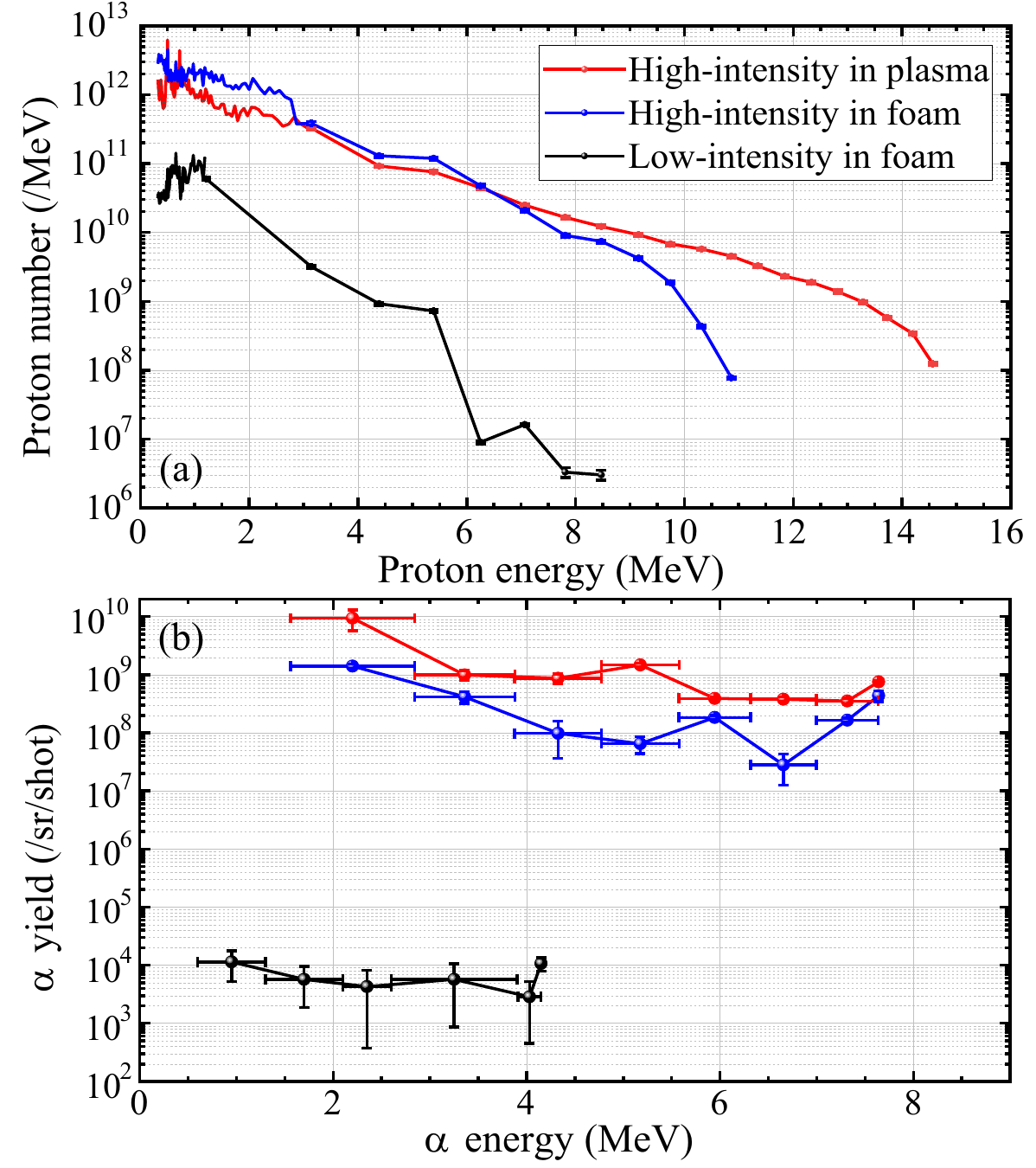}
\end{minipage}
\caption{The measured energy spectra of (a) proton beams and (b) $\alpha$-particle beams from plasma (red curve) and foam targets with protons in high intensity (blue curve)  and low intensity (black curve), respectively. The proton energy spectra are measured by the RCF stack (line with symbol) combined with the TPS (only line, low-energy part). The energy spectra of $\alpha$ particles were obtained from CR39 detectors with Al foils in different thicknesses located along $52^{\circ}$ direction. The error bars represent the level of uncertainty in the calibration of the RCF and the statistical uncertainties in $\alpha$-particle yield, respectively. 
\label{fig2}}
\end{figure}

\section{Results and Discussion}

The energy spectra of the incident proton beam measured by the RCF stack for the high energy part, combined with TPS measurements for the low energy part are shown in Fig.~\ref{fig2} (a) for three cases, presenting a broad exponential distribution. Thus, the absolute intensity $N_p$ of the incident protons interacting with plasma or foam targets can be obtained approximately. For several shots, the ps laser energy was lower by about 40~J, reducing the total proton intensity and energy (black curve). The energy spectra of the generated $\alpha$ particles were obtained by the standard technique from CR39 meaurements with Al filters \cite{SM2022} as shown in Fig.~\ref{fig2} (b). Note that the cutoff energies for $\alpha$ particles for high and low-intensity cases are 7.6~MeV and 4.1~MeV due to the thickest filter of 40~$\mathrm{\mu m}$ and 17~$\mathrm{\mu m}$, respectively.

The $\alpha$-particle yield  from the cold foam target for the low-intensity case is on the order of $10^{4}$/sr/shot, as shown in Fig.~\ref{fig2}  in black on the bottom.  By increasing the proton beam intensity, the $\alpha$-particle yield is nonlinearly enhanced both for the foam and plasma cases, as shown in Fig.~\ref{fig2} in blue and red, respectively.  

\begin{table}[h]
\caption{$\alpha$-particle yield per steradian from CR39 detectors located along $52^{\circ}$ direction. Total proton number $N_p$, $\alpha$-particle yield $N_{\alpha}$ (/sr) and $\alpha$-particle yield per proton $N_{\alpha}/N_p$ (/$p$/sr) for the cases of plasma and cold foam in high-intensity (HI) and low-intensity (LI) conditions, respectively.  The fusion reaction probalility is represented as $P$, where $P=\frac{4\pi}{3} N_{\alpha}/N_p$. }
\label{Tab:yield}
\begin{center}
\renewcommand\arraystretch{1.2}
\begin{tabular}{c|c|c|c|c}
  \hline\hline
  \multirow{2}{*}{Cases} & HI- & HI- & LI- & Beam-target \\ [0.01 ex]
  &plasma &foam &foam &prediction\\ [0.01 ex]
  \hline
 \multirow{2}{*}{$N_p$} & $2.7\times10^{12}$ & $4.8\times10^{12}$ & $1.3\times10^{11}$  & \multirow{2}{*}{$1.3\times10^{11}$}  \\ [0.01 ex]
  & $\pm9.5\%$ & $\pm12.7\%$ & $\pm7.4\%$ &\\ [0.01 ex]
 \hline
  $N_{\alpha}$ & $(1.5\pm0.4)$ & $(2.8\pm0.2)$ & $(4.1\pm1.0)$ & \multirow{2}{*}{$2.2\times10^{4}$} \\ [0.01 ex]
  [/sr] & $\times10^{10}$ & $\times10^{9}$ & $\times10^{4}$ & \\ [0.01 ex]
 \hline
  $N_{\alpha}/N_p$ & $(5.6\pm1.4)$ & $(5.8\pm0.4)$ & $(3\pm0.8)$ & \multirow{2}{*}{$1.7\times10^{-7}$} \\ [0.01 ex]
  [/$p$/sr] & $\times 10^{-3}$ & $\times10^{-4}$ & $\times 10^{-7}$ & \\ [0.01 ex]
\hline
  $P$ & $2.3\times10^{-2}$ & $2.4\times10^{-3}$ & $1.3\times10^{-6}$ & $7.3\times10^{-7}$ \\
 \hline\hline
\end{tabular}
\end{center}
\end{table}

Table~\ref{Tab:yield} summarizes the $\alpha$-particle yield $N_{\alpha}$, the yield normalized to the proton intensity $N_{\alpha}/N_p$ and the fusion reaction probalility $P$ for the above three cases. For comparison, the beam-target prediction of p$^{11}$B fusion reaction is provided as well \cite{SM2022}. Using the same parameters, we see good agreement with the low-intensity case. As conclusion of Table~\ref{Tab:yield}, we observe an increase of 4 orders of magnitude of the normalized  $\alpha$-particle yield, and the fusion reaction probalility rises up to $2.3\times10^{-2}$. That means every 43th proton will induce one p$^{11}$B fusion event.

\begin{figure}
\centering
\begin{minipage}[b]{0.5\textwidth}
\centering
\includegraphics[width=3.4in]{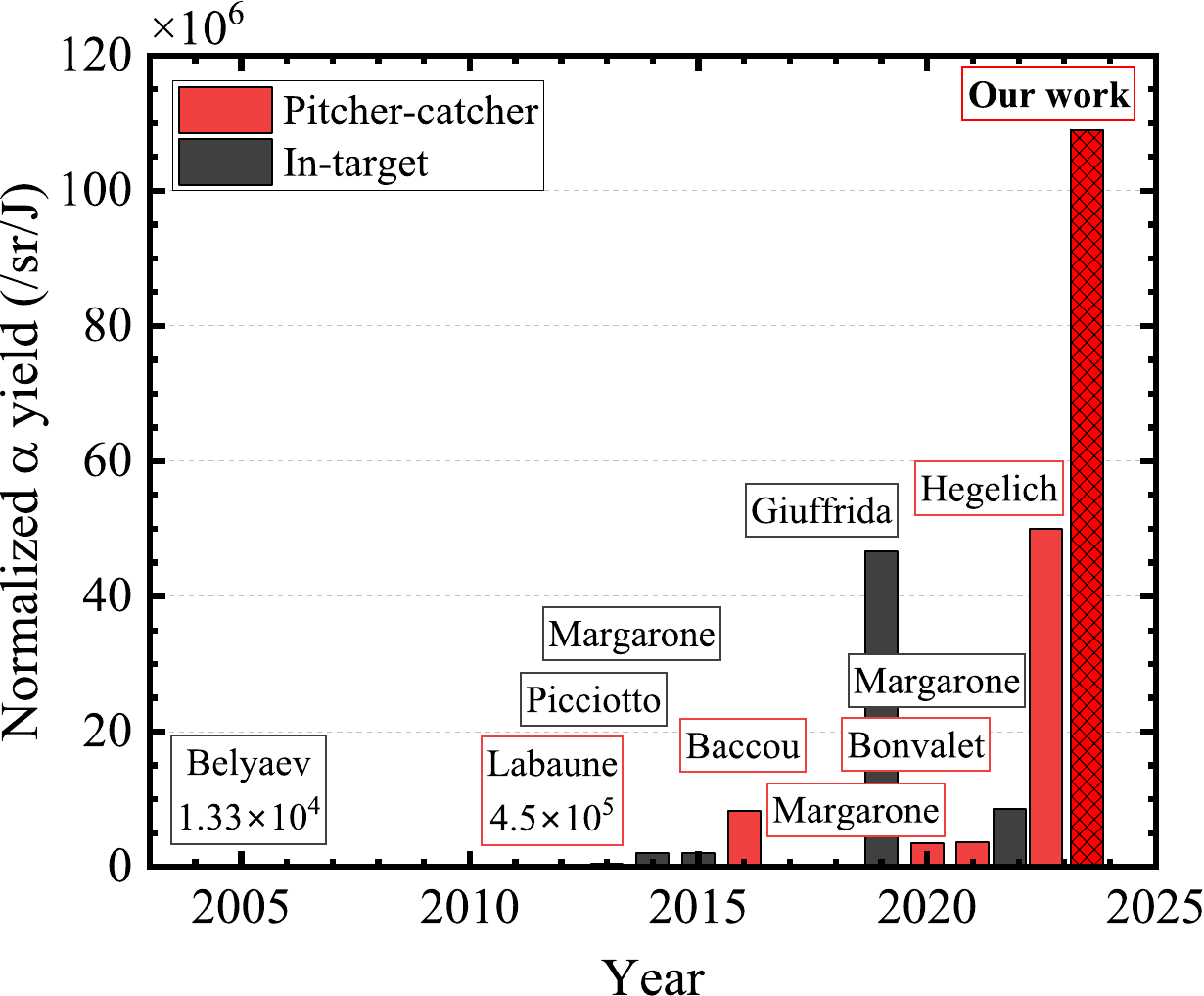}
\end{minipage}
\caption{The laser-driven p$^{11}$B fusion progress in maximum $\alpha-$particle yield normalized to the laser energy delivered on target for both ``in-target" (black) \cite{Belyaev2005,Kimura2009,Giuffrida2020,Picciotto2014,Margarone2015,Margarone2022} 
and ``pitcher-catcher" (red) \cite{Bonvalet2021,Margarone2020g,Baccou2015n,Labaune2013} irradiation geometries. Red bar with grid pattern denotes our results. 
\label{fig3}}
\end{figure}

In Fig.~\ref{fig3}, we analyzed the current status of the laser-driven p$^{11}$B fusion. Two different schemes exist. In the ``pitcher-catcher" scheme, the laser generates the proton beam and subsequently induce the fusion reaction. While in the ``in-target" scheme, the laser irradiates the boron target directly. We normalized the $\alpha$-particle yield to the laser energy for both schemes as shown in Fig.~\ref{fig3}. Compared to other impressive works \cite{Mehlhorn2022} (see Fig.~\ref{fig3}), the normalized $\alpha$-particle yield in our experiment is the highest at present. This value can be higher if all the accelerated protons are used. The total energy of the generated $\alpha$ particles is about 0.1~J. If we consider the total  energy of the incident proton beam ($\sim0.85~$J), the energy conversion efficiency from the incident protons to the $\alpha$ particles is as high as $12\%$.

Even though a full understanding of the details is not yet reached, there are a number of facts that qualitatively may explain the observed effects. In a previous experiment \cite{Ren2020}, we observed a high degree of stopping for laser-accelerated intense proton beams in dense ionized matter similar to the conditions of this experiment. The enhanced stopping in that case was attributed to collective electromagnetic effects, resulting in strong electric fields on the order of $10^{9}$ V/m. In the current situation, the field strength is even higher. Recent simulation shows that the electric fields induced by high-current charged particle beams in foam are much higher than those in homogenous materials \cite{Jiangk2023}. Also bulk acceleration as discussed in Ref.\cite{Zhangj2005}, contributes to the acceleration of the local background ions, so that the plasma ions do no longer obey the classical Maxwellian distribution. This suprathermal condition is sufficient to induce p$^{11}$B fusion reactions \cite{Hartouni2022, Mehlhorn2022,Rider1997}. As a result, we observed the fusion probalility $P$ nonlinearly beyond the prediction of the beam-target reaction.

As reported previously, the 3D sponge-like structure of the foam target with a pore size around 1~$\mu$m can have an effect on the fusion yield via improving the ion-ion collisions \cite{Zhangyh2019}. In literature it is also discussed that up-scattering via elastic collisions of background protons with the fusion fast $\alpha$ particles can also increase the high-energy proton population, leading to the enhancement of fusion yield \cite{Hartouni2022}.  Strong electric fields may also affect the fusion yield by increasing the tunneling probability \cite{Lvwj2022}. In addition, high-energy protons with broad spectra would open more inelastic scattering and additional reaction channels with the backgroud carbon- and oxygen nuclei \cite{Barnard1966,Gruhle1977,Sajjad1986}, but our opinion this can be neglected. All effects mentioned here contribute to an increased $\alpha$ yield, but more modelling of the conditions and high-precision experiments are necessary to quantitatively understand the results. It is of paramount importance to study the non-equilibrum thermonuclear fusion scenarios in such complex and extreme environments. 

In conclusion, we show a significant enhancement of the p$^{11}$B fusion yield per joule compared to previously published results. The normalized yield is up to four orders of magnitude beyond the classical beam-target prediction. This is mainly due to the nonlinear increase of the yield with the proton beam intensity. The foam structure, in addition, the plasma state, plays an important role to create the high-intensity electric fields and suprathermal conditions. Our method provides a new pathway to explore the aneutronic fusion reaction in dense non-equilibrium plasma environment induced by the laser-accelerated high-intensity proton beams. This may well turn the p$^{11}$B fusion from a competitor in the fusion race with no chance to win (also-ran) into a competitor who is catching up (comer).  
 
{\it Acknowledgement}: We sincerely thank the staff from Laser Fusion Research Center, Mianyang for the laser system running and target fabrication. This work has been supported by the National Key Research and Development Program of China (Grant No. 2019YFA0404900, 2022YFA1603302), National Natural Science Foundation of China (Grant Nos. 12105218, U2030104,  U2230112, 12120101005 and 12175174), Chinese Science Challenge Project (Grant No. TZ2016005), China Postdoctoral Science Foundation (Grant No. 2020M673367), Science and Technology on Plasma Physics Laboratory (Grant No. J202108010) and the Fundamental Research Funds for the Central Universities.

{\it Author contributions}: Yongtao Zhao conceived this work and organized the experiments with Weimin Zhou and Jieru Ren. Wenqing Wei, Zhigang Deng, Jieru Ren and Yongtao Zhao carried out the experiment together with the high-power laser team (Zongqing Zhao, Weimin Zhou, and Yuqiu Gu), the plasma diagnostics team (Quanping Fan, Shaoyi Wang, Bubo Ma, Bo Cui, Rui Cheng, Zhao Wang, Yu Lei, and Leifeng Cao), ion beam characterization team (Lirong Liu, Fangfang Li, Zhongmin Hu, Benzheng Chen, and Weiwu Wang), $\alpha$ particle characterization team (Wei Qi, Hao Xu, Xing Xu, Shizheng Zhang, and Jian Teng), target operation team (Feng Lu, Lei Yang, Guanchao Zhao, and Bing Liu). Wenqing Wei, Shizheng Zhang, Zhigang Deng, Hao Xu, and Lirong Liu analyzed the main part of the experimental data. Wenqing Wei, Shizheng Zhang, Jialin Zhang, Jieru Ren, and Yongtao Zhao performed the data analysis and Geant4 simulations. Yuqiu Gu, Bing Liu, Minsheng Liu, Huasheng Xie, Jianxing Li, Xueguang Ren, Zhongfeng Xu, Guoqing Xiao, Hongwei Zhao, and Leifeng Cao contribute in the physical discussion. Wenqing Wei, Dieter H.H. Hoffmann, and Yongtao Zhao wrote the paper.






\bibliography{reference}

\end{document}